\documentstyle[prl,aps,multicol,epsf]{revtex}
\begin{document}
\draft
\title{Spin polarization of strongly interacting 2D electrons: the role of
disorder.}
\author{S. A. Vitkalov and M.P. Sarachik}
\address{Physics Department, City College of the City
University of New York, New York, New York 10031}
\author{T.~M.~Klapwijk}
\address{Delft University of Technology, Department of Applied Physics,
2628 CJ Delft, The Netherlands}
\date{\today}
\maketitle

\begin{abstract}
In high-mobility silicon MOSFET's, the $g^*m^*$ inferred indirectly from
magnetoconductance and magnetoresistance measurements with the assumption
that $g^*\mu_BH_s=2E_F$ are in
surprisingly good agreement with $g^*m^*$ obtained by direct measurement
of Shubnikov-de Haas oscillations.  The enhanced
susceptibility $\chi^* \propto (g^*m^*)$ exhibits critical behavior
of the form $\chi^* \propto (n - n_0)^{-\alpha}$.  We examine the
significance of the field scale $H_s$ derived from transport
measurements, and show that this field signals the onset of full spin
polarization only in the absence of disorder.  Our results suggest that
disorder becomes increasingly important as the electron density is
reduced toward the transition.  

\end{abstract}


\begin{multicols}{2}

Two dimensional systems of electrons \cite{krav,papadakis,hanein2}
and holes \cite{coleridge,hanein,simmons} have been the focus of a great
deal of attention during the last few years.  In contrast with
expectations for noninteracting \cite{gang} or weakly interacting
\cite{altshuler} electrons in two dimensions, these strongly
interacting systems exhibit metallic behavior in the absence of a magnetic
field: above some characteristic electron (hole) density, $n_c$, their
resistivities decrease with decreasing temperature.  Whether there is a
genuine metallic phase and a true metal-insulator transition in these
materials continues to be the subject of lively debate \cite {AKS}.

Experimental results have been obtained in the 2D system of
electrons in silicon MOSFET's that indicate that the response to a
magnetic field applied in the plane of the electrons increases
dramatically as the electron density is decreased toward $n_c$.  Based on
a study of the scaled magnetoconductance as a function of temperature and
electron density, Vitkalov {\it et al.}
\cite{vitkalov_ferro} have identified an energy scale $\Delta$ that
decreases with decreasing density and extrapolates to zero in the limit $T
\rightarrow 0$ at a density $n_0$ in the vicinity of $n_c$; this was
interpreted as evidence of a quantum phase transition at $n_0$.  
From studies at very low temperatures of the magnetoresistance as a
function of electron density, Shashkin {\it et al.} \cite{Shashkin}
inferred that the two-dimensional system of electrons in silicon
inversion layers approaches a ferromagnetic instability at the critical
density $n_c$ for the zero-field metal-insulator transition.  From a
determination of the enhanced spin susceptibility derived from
Shubnikov-de Haas measurements down to low densities, Pudalov
{\it et al.} \cite{{pudalov_SdH}} have claimed there is no spontaneous
spin polarization for electron densities above $n = 8.34 \times 10^{10}$
cm$^{-2} \approx n_c$, although they could not exclude this for lower
densities.  The possibility that a magnetically ordered phase exists in
the limit $T \rightarrow 0$ in dilute two-dimensional silicon inversion
layers is intriguing and bears further investigation.

In this paper we show that there is very good agreement
between values reported for $g^*m^*$ as a function of electron density in
high-mobility silicon MOSFET's obtained directly from measurements of the
Shubnikov-de Haas oscillations \cite{{pudalov_SdH}} and those inferred
indirectly from magnetoconductance and magnetoresistance measurements by
two different groups using different methods of analysis and the
assumption that
$g^*\mu_BH_s=2E_F$ \cite{vitkalov_ferro,Shashkin,comparison}.  Here $g^*$
is the enhanced $g$-factor, $m^*$ is the enhanced electron mass,
$\mu_B$ is Boltzmann's factor, $E_F$ is the Fermi energy, and $H_s$ is a
characteristic field scale determined by different methods from in-plane
magnetoconductance \cite{vitkalov_ferro} and magnetoresistance \cite
{Shashkin} experiments.  The enhanced susceptibility $\chi^* \propto
(g^*m^*)$ exhibits critical behavior of the form $\chi^* \propto (n -
n_0)^{-\alpha}$.  Data from the three experimental groups yield exponents
$\alpha$ of $0.23, 0.24$ and $0.27$, and critical densities between
$0.88$ and $1.04 \times 10^{11}$ cm$^{-2}$.  We examine the significance
of the field scale $H_s$, and show that this field signals the onset
of full spin polarization only in the absence of disorder.  Our results
suggest that disorder becomes increasingly important as the electron
density is reduced toward the transition.  

Measurements of Shubnikov-de Haas oscillations in high-mobility silicon
MOSFET's with high electron densities have shown that the magnetic field
required to achieve complete polarization of the electron spins is
approximately the same as that required to saturate the magnetoresistance
to a constant value \cite{okamoto,pol1,pol2}.  For the relatively high
densities used in these experiments, the field $H_\rho$ correponding to
saturation of the magnetoresistance is approximately the same as the
field $H_\sigma$ above which there is apparent saturation of the
magnetoconductance.  As we show below, this equivalence  breaks down at
lower densities.  A clear example is illustrated in Fig. \ref{rhosigma},
where the resistivity and conductivity are shown as a function of
in-plane magnetic field for a silicon MOSFET with electron density near
the critical density, $n_c$, for the metal-insulator transition.  The
saturation field $H_\rho$ derived from the resistivity is considerably
larger than the field $H_\sigma$ above which the conductivity saturates. 
This can be understood with reference to the band diagrams shown as
insets to Fig. \ref{rhosigma}.  In the absence of disorder, \vbox{
\vspace{0.1 in}
\hbox{
\hspace{-0.4in} 
\epsfxsize 3.6 in \epsfbox{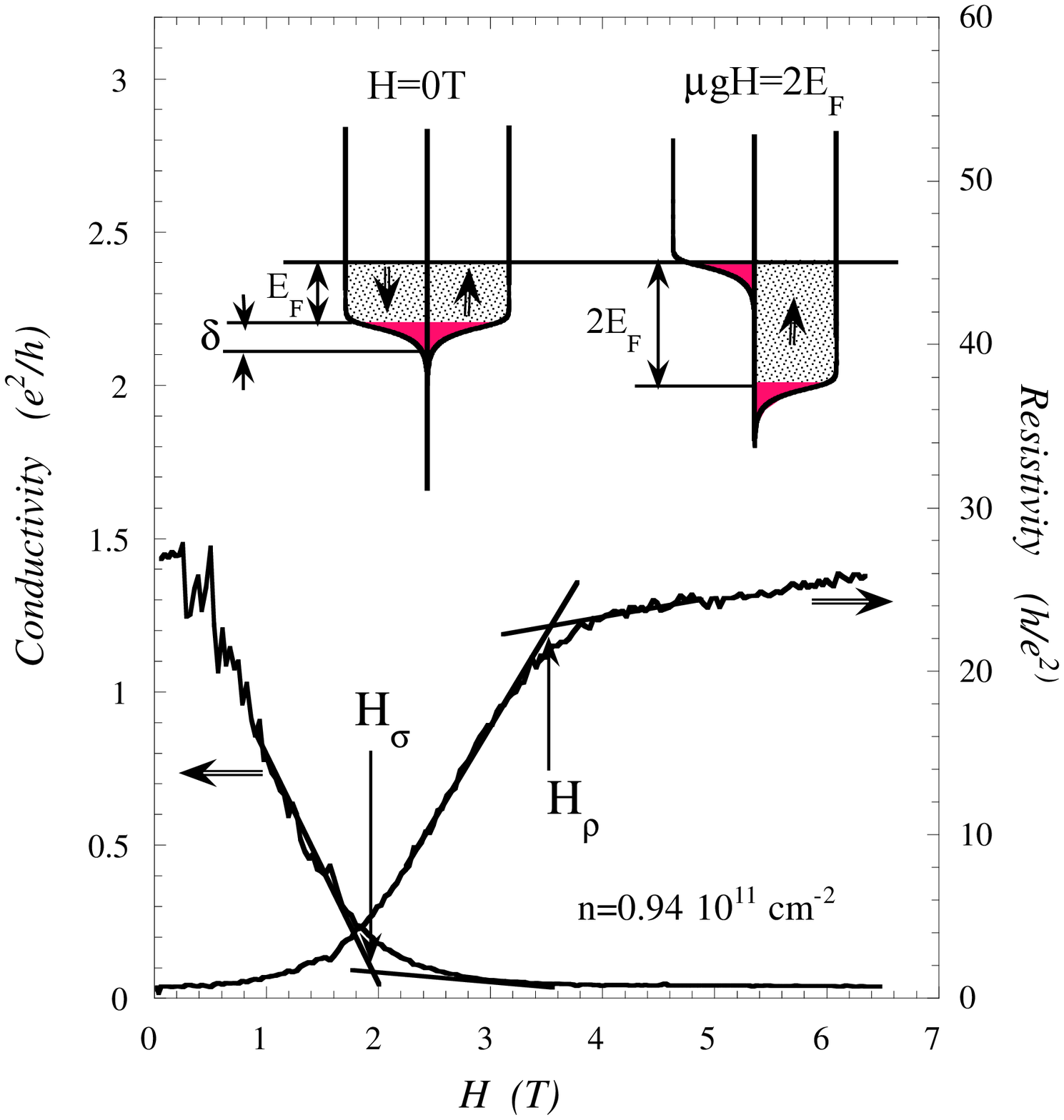} 
}
}
\vskip -0.7cm
\refstepcounter{figure}
\parbox[b]{2.9in}{\baselineskip=12pt FIG.~\thefigure.
For a silicon MOSFET with electron density $0.94 \times 10^{11}$
cm$^{-2}$, the conductivity (left curve) and resistivity (right curve)
are shown as a function of in-plane magnetic field at temperature
$T=0.26$ K; the saturation fields deduced from the resistivity and the
conductivity are labeled $H_\rho$ and $H_\sigma$, respectively.  The
insets show schematic diagrams of the electron bands (see
text for discussion).

\vspace{0.3in}
}
\label{rhosigma}

\noindent all electron
states are extended, band-tailing plays a negligible role, and full spin
polarization is achieved when the Zeeman energy is sufficient to
completely depopulate the minority spin band: 
\begin{equation}
g^* \mu_B H_{band} = 2 E_F;
\label{band}
\end{equation}
here $g^*$ is the enhanced $g$-factor, $\mu_B$ is Boltzmann's
factor, $H_{band}$ is the magnetic field required to fully polarize the
system in the absence of disorder, and $E_F$ is the Fermi energy. 
Disorder is weak at high electron densities and one expects $H_{band}
\approx H_\rho
\approx H_\sigma$.  

As the density is decreased and disorder and the band-tails
become more important, complete spin alignment requires the
application of a larger magnetic field to fully polarize the tail states
as well as the extended states:
\begin{equation}
g^* \mu_B H_{tail+band} = 2 E_F + \delta
\label{tail}
\end{equation}
where we've assumed the band tail has an effective energy
width $\delta$ \cite{strongdisorder}.  

Except very near the transition, the number of states in the
band tails in the case of samples of reasonably high mobility is much
smaller than the number of extended states; at the same time, the energy
width $\delta$ becomes appreciable as the density decreases and the
disorder increases.  The field required to align the electrons in the
higher mobility band states can thus differ substantially from the
magnetic field needed to polarize
$all$ the electrons
\cite{pudalov2}.  While the (small number) of  tail states make a minor
contribution to the conductivity, the resistivity is considerably more
sensitive to the low-mobility states in the tail of the distribution, and
consequently
$H_\rho > H_\sigma$ as is evident in Fig.\ref{rhosigma}.  We suggest that
$H_\sigma \approx H_{band}$ and
$H_\rho \approx H_{tail+band}$.

\vbox{
\vspace{0.3 in}
\hbox{
\hspace{-0.1in} 
\epsfxsize 3.4 in \epsfbox{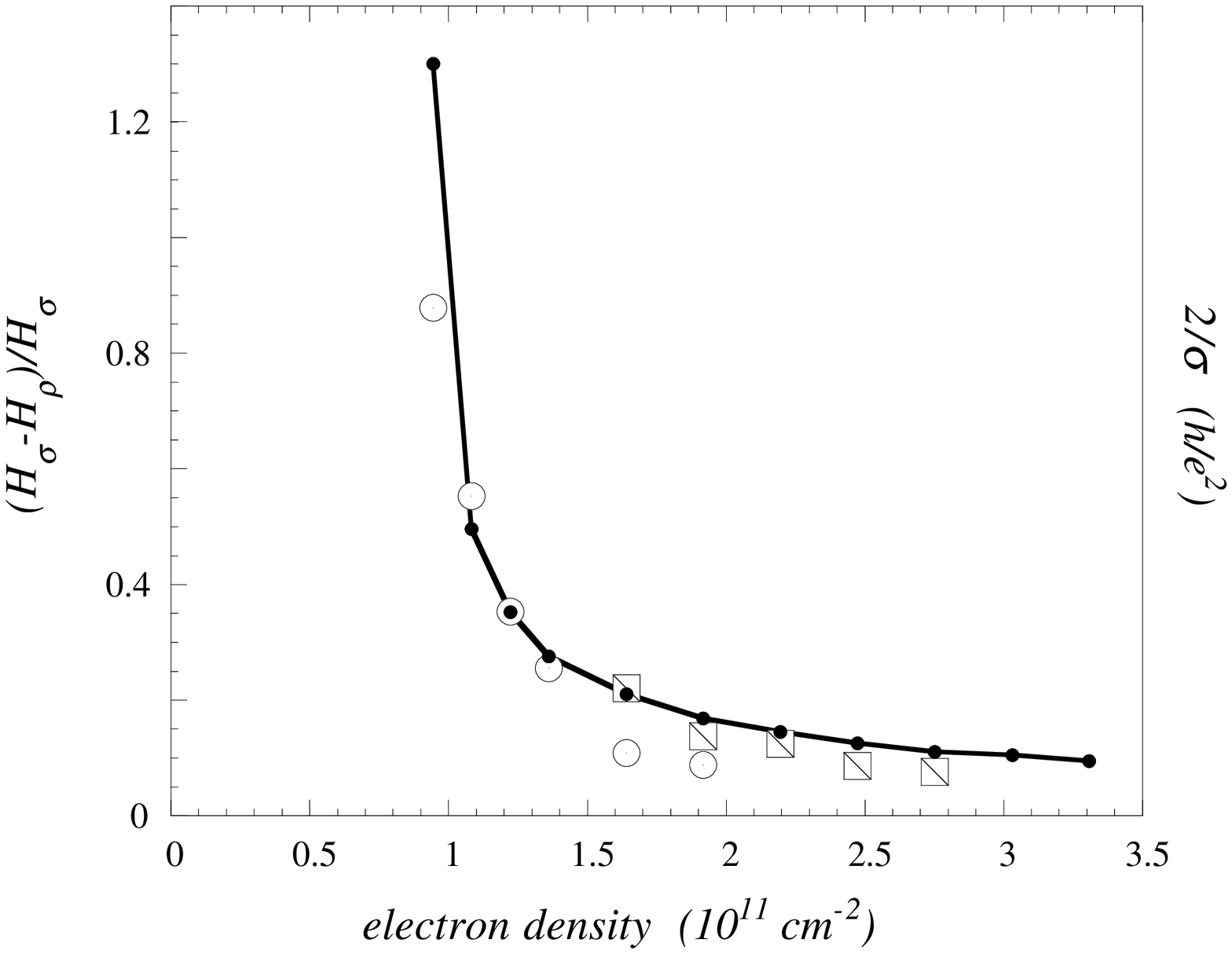} 
}
}
\vskip 0.5cm
\refstepcounter{figure}
\parbox[b]{3.3in}{\baselineskip=12pt FIG.~\thefigure.
The fractional difference $(H_\rho - H_\sigma)/H_\sigma$ (open symbols)
and $2/\sigma$ (closed symbols) versus electron density; $H_\rho$ and
$H_\sigma$ are the saturation fields deduced from resistivity and
conductivity curves, respectively.  The open circles and closed symbols
refer to data taken at $0.26$ K.  The open squares are data obtained at
$0.1$ K on a different MOSFET. 

\vspace{0.10in}
}
\label{difference}

The fractional difference between $H_\rho$ and $H_\sigma$, $\Delta H/H=
(H_\rho - H_\sigma)/H_\sigma$, is shown as a function of electron density
in Fig. \ref{difference}; $\Delta H/H$ increases rapidly with decreasing
electron density when disorder becomes more dominant.  The quantity
$2/\sigma$ is plotted for comparison through the following argument.  For
weak scattering, the parameter $\delta$ is on the order of the scattering
rate: $\delta \sim \hbar/\tau$. With Eqs. \ref{band} and \ref{tail},
this gives $\Delta H/H =\delta/2E_F=\hbar/2(E_F \tau)$.   Using the
expression for the Drude conductivity $\sigma = ne^2\tau/m^*$, and the
Fermi energy $E_F/\hbar = (nh)/g_vg_sm^*$ with a valley degeneracy $g_v=2$
and spin degeneracy
$g_s=2$, one obtains $\Delta H/H=(e^2/h)(2/\sigma)$.  The correlation
between 
$\Delta H/H$ and $2/\sigma$ is evident in Fig. \ref{difference}.

In an earlier paper \cite{vitkalov_ferro}, we showed that the
magnetoconductance of silicon MOSFET's can be scaled onto a single
curve by plotting $[\sigma(H) - \sigma(0)]/[\sigma(H=\infty) -
\sigma(0)]$ as a function of $H/H_s$.  The parameter $H_s$ obtained by
this method is proportional to $H_\sigma$ discussed above.  For high 
densities where disorder plays a small role, the magnetic field
$H_\sigma$ needed to saturate the conductivity is very nearly equal to
the field required to obtain full spin polarization.  At lower densities,
the saturation fields deduced from the resistivity and the conductivity
are not the same, and we have argued that the difference is associated
with the effect of electrons in the states in the band tails. 
We've suggested that $H_\sigma$ is the magnetic field required to polarize
the band states; the Zeeman energy and $g^*m^*$ are then given by Eq.
\ref{band} with $H_{band} = H_\sigma$.  The tail states remain unpolarized
in $H=H_\sigma$.  However, except perhaps very near the transition (or in
samples of very low mobility), they represent a small fraction of the
electrons, so that the system is close to full spin polarization.

\vbox{
\vspace{0.4 in}
\hbox{
\hspace{-0.2in} 
\epsfxsize 3.4 in \epsfbox{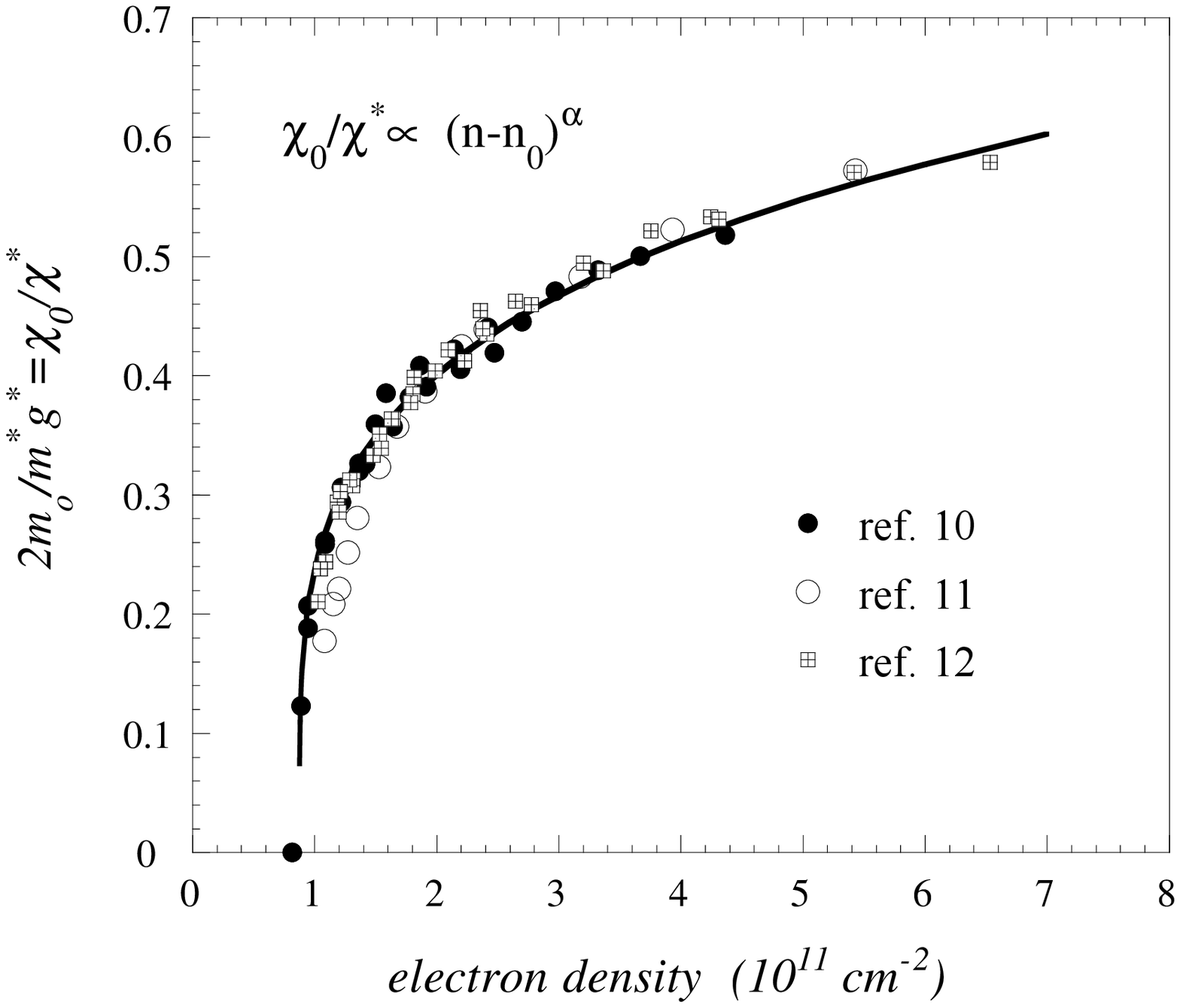} 
}
}
\vskip 0.5cm
\refstepcounter{figure}
\parbox[b]{3.1in}{\baselineskip=12pt FIG.~\thefigure.
The inverse of the enhanced susceptibility $\chi_0/\chi^*$ versus electron
density obtained by Vitkalov {\it et al.} \cite{vitkalov_ferro}, Shashkin
{\it et al.}
\cite{Shashkin}, and Pudalov {\it et al.} \cite{pudalov_SdH}.  Data are
normalized to the Shubnikov-de Haas values at high densities.  The curve
is a fit to the critical form
$\chi_0/\chi^* = A(n-n_0)^\alpha$ for the data of ref.
\cite{vitkalov_ferro} (excluding the point shown at $\chi_0/\chi^* = 0$).

\vspace{0.10in}
}
\label{critical}

Fig. \ref{critical} shows $2m_0/m^*g^* = \chi_0/\chi^*$ as a function of
electron density $n_s$ obtained from our data \cite{vitkalov_ferro},
by Sashkin {\it et al.} \cite{Shashkin}, and Pudalov {\it et al.}
\cite{pudalov_SdH}.  Here
$\chi^*/\chi_0$ is the enhanced susceptibility normalized to its free
electron value, and
$\chi_0/\chi^*$ is its inverse.  The closed circles denote values obtained
from scaling our data for the in-plane magnetoconductance and the
assumption that $g^*
\mu_B H_\sigma = 2 E_F$; the open circles were obtained by Shashkin {\it
et al.}
\cite{Shashkin}from magnetoresistance measurements using a different
data-fitting procedure and the same assumption as above; the squares
are from direct Shubnikov-de Haas measurements of Pudalov {\it et
al.}\cite{pudalov_SdH}.  The data of Shashkin {\it et al.} decrease
somewhat more rapidly at low densities than the others.  However, the
three sets obtained by different groups using different measurements and
different methods of analysis agree surprisingly well.  Again, this
indicates that the small number of states in the band tails in
high-mobility MOSFET's play a neglibible role.  A fit to the critical form
\begin{equation}
\chi_0/\chi^* \propto (n-n_0)^\alpha,
\label{crit}
\end{equation}
yields the following values for the three data sets considered: for the
Shubnikov-de Haas data of Pudalov {\it et al.} \cite{pudalov_SdH}
$\alpha = 0.23$, $n_0=0.96 \times 10^{11}$ cm$^{-2}$; for the
magnetoconductance data of Shashkin {\it et al.} \cite{Shashkin}
$\alpha = 0.27$,
$n_0=1.04 \times 10^{11}$ cm$^{-2}$; and for our data
\cite{vitkalov_ferro} $\alpha = 0.24$, $n_0=0.88 \times
10^{11}$ cm$^{-2}$.

We have argued above that for high-mobility samples, the difference
$(H_\rho - H_\sigma)$ is associated with the effect of a small fraction
of the electrons  in the band tails.  The characteristic field $H_s$
obtained in our earlier work was determined from scaling the
magnetoconductance, which is a measure of the field required to align the
band states while leaving a few electrons in the tail states
unpolarized.  Shashkin {\it et al.} determined a field scale by matching
magnetoresistance data at $low$ magnetic fields; close examination shows
that this procedure does not produce a match at high fields (note that
their data is shown on a logarithmic scale, which deemphasizes
differences between the curves at high values of magnetic field).  Both
methods are sensitive to the contribution of the extended state and
minimize the effect of the states in the band tails.  These procedures
yield reliable measures for the behavior of the system at high
electron densities where disorder does not play an important role.  This
accounts for the surprisingly good agreement between the $g^*m^*$
obtained from transport experiments and those found by direct measurement
of the Shubnikov-de Haas oscillations.  At densities very near the
transition (and for very low mobility MOSFET's) one should expect this
correspondence to break down as disorder becomes more dominant.  We
suggest that an understanding of any phase transition that occurs in this
regime must incorporate the effect of disorder in a central way.

We thank M. Gershenson and S. V. Kravchenko for providing data
for Fig. \ref{critical}.  This work was supported by the US Department of
Energy under Grant No.~DE-FG02-84ER45153.

\end{multicols}


\begin{references}

\bibitem{krav} S.\ V.\ Kravchenko, G.\ V.\ Kravchenko, J.\ E.\
Furneaux, V.\ M.\ Pudalov, \and M.\ D'Iorio, Phys.\ Rev.\ B
{\bf 50}, 8039 (1994); S.\ V.\ Kravchenko, W.\ E.\ Mason, G.\ E.\
Bowker, J.\ E.\ Furneaux, V.\ M.\ Pudalov, \and M.\ D'Iorio, 
Phys.\ Rev.\ B {\bf 51}, 7038 (1995); S. V. Kravchenko, D.
Simonian, M. P. Sarachik, Whitney Mason, \and J. E. Furneaux,
Phys.\ Rev.\ Lett. {\bf 77}, 4938 (1996).

\bibitem{papadakis} S. J. Papadakis and M. Shayegan, Phys. Rev.
B {\bf 57}, R15068 (1998).

\bibitem{hanein2} Y. Hanein, D. Shahar, J. Yoon, C. C. Li, D. C. Tsui,
\and H. Shtrikman, Phys. Rev. B {\bf 58}, R13338 (1998).

\bibitem{coleridge} P.~T.~Coleridge, R.~L.~Williams, Y.~Feng, \and
P.~Zawadzki, Phys.\ Rev.\ B {\bf 56}, R12764 (1997).

\bibitem{hanein} Y.~Hanein, U. Meirav, D. Shahar, C. C. Li, D. C. Tsui, 
\and H. Shtrikman, Phys.\ Rev.\ Lett. {\bf 80} 1288 (1998).

\bibitem{simmons} M.~Y.~Simmons, A. R. Hamilton, M. Pepper, E. H.
Linfield, P. D. Rose, D. A. Ritchie, A. K. Savchenko, \and T. G.
Griffiths, Phys.\ Rev.\ Lett. {\bf 80} 1292 (1998).

\bibitem{gang} E. Abrahams, P. W. Anderson, D. C. Licciardello, and
T. V. Ramakrishnan. Phys. Rev. Lett. {\bf 42} 673 (1979).

\bibitem{altshuler} B. L. Altshuler, A. G. Aronov, and P. A. Lee,
Phys. Rev. Lett. {\bf 44}, 1288 (1980).

\bibitem{AKS} For a review see E. Abrahams, S. V. Kravchenko, and M. P.
Sarachik, Rev. Mod. Phys.{\bf 73}, 251 (2001).

\bibitem{vitkalov_ferro} S. A. Vitkalov, H. Zheng, K. M. Mertes, M. P.
Sarachik and T. M. Klapwijk, Phys. Rev. Lett. {\bf 87}, 086401, (2001).

\bibitem{Shashkin} A. A. Shashkin, S. V. Kravchenko, V. T. Dolgopolov,
and T. M. Klapwijk, Phys. Rev. Lett. {\bf 87}, 086801, (2001).

\bibitem{pudalov_SdH}  V. M. Pudalov, M. Gershenson, H. Kojima, preprint
cond-mat/0110160 (2001).

\bibitem{comparison}  A similar comparison was presented in a comment by
S. V. Kravchenko, preprint cond-mat/0106056 (2001).

\bibitem{strongdisorder}  For low disorder, one expects that the density
of states is much smaller in the band tail than for the extended states,
and small variations of $E_F$ can be neglected in Eq. \ref{tail}.

\bibitem{pudalov2}  The possible role of
localized or bound states was considered by V. M. Pudalov, G. Brunthaler,
A. Prinz, and G. Bauer, to be published in Phys. Rev. Lett. (2002);
preprint cond-mat/0004206 (2000).

\bibitem{okamoto} T. Okamoto, K. Hosoya, S. Kawaji, and A. Yagi, Phys.
Rev.Lett. {\bf 82}, 3875 (1999).

\bibitem{pol1} S. A. Vitkalov, H. Zheng, K. M. Mertes, M. P. Sarachik
and T. M. Klapwijk, Phys. Rev. Lett. {\bf 85}, 2164 (2000).

\bibitem{pol2} S. A. Vitkalov, M. P. Sarachik, and T. M.
Klapwijk,  Phys. Rev. {\bf B} {\bf 64}, 073101, (2001).

\end{references}
\end{document}